\documentclass[twocolumn]{aastex631}

\usepackage{amsmath}
\usepackage{cleveref}

\begin{document}

\title{The Radio Counterpart to the Fast X-ray Transient EP240414a}

\correspondingauthor{Joe S. Bright}
\email{joe.bright@physics.ox.ac.uk}

\author[0000-0002-7735-5796]{Joe S. Bright}
\affiliation{Astrophysics, Department of Physics, The University of
Oxford, Keble Road, Oxford, OX1 3RH, UK}
\affiliation{Breakthrough Listen, Astrophysics, Department of Physics, The University of Oxford, Keble Road, Oxford, OX1 3RH, UK}

\author[0000-0002-0426-3276]{Francesco Carotenuto}
\affiliation{Astrophysics, Department of Physics, The University of Oxford, Keble Road, Oxford, OX1 3RH, UK}

\author{Rob Fender}
\affiliation{Astrophysics, Department of Physics, The University of Oxford, Keble Road, Oxford, OX1 3RH, UK}
\affiliation{Department of Astronomy, University of Cape Town, Private Bag X3, Rondebosch 7701, South Africa}

\author[0009-0008-0662-1293]{Carmen Choza}
\affiliation{SETI Institute, 339 Bernardo Ave, Suite 200, Mountain View, CA 94043, USA}

\author{Andrew Mummery}
\affiliation{Oxford Theoretical Physics, Beecroft Building, Clarendon Laboratory, Parks Road, Oxford, OX1 3PU, UK}

\author[0000-0001-5679-0695]{Peter G. Jonker}
\affiliation{Department of Astrophysics/IMAPP, Radboud University, P.O. Box 9010, 6500 GL, Nĳmegen, The Netherlands}
\affiliation{SRON, Netherlands Institute for Space Research, Niels Bohrweg 4, 2333 CA Leiden, The Netherlands}

\author[0000-0002-8229-1731]{Stephen J. Smartt}
\affiliation{Astrophysics, Department of Physics, The University of Oxford, Keble Road, Oxford, OX1 3RH, UK}
\affiliation{Astrophysics Research Centre, School of Mathematics and Physics, Queen’s University Belfast, BT7 1NN, UK}

% alphabetical

\author[0000-0003-3197-2294]{David R. DeBoer}
\affiliation{Radio Astronomy Lab, University of California, Berkeley, CA, USA}

\author[0000-0002-0161-7243]{Wael Farah}
\affiliation{SETI Institute, 339 Bernardo Ave, Suite 200, Mountain View, CA 94043, USA}
\affiliation{Berkeley SETI Research Centre, University of California, Berkeley, CA 94720, USA}

\author[0000-0002-3493-7737]{James Matthews}
\affiliation{Astrophysics, Department of Physics, The University of Oxford, Keble Road, Oxford, OX1 3RH, UK}

\author[0000-0002-3430-7671]{Alexander W. Pollak}
\affiliation{SETI Institute, 339 Bernardo Ave, Suite 200, Mountain View, CA 94043, USA}

\author[0000-0003-2705-4941]{Lauren Rhodes}
\affiliation{Astrophysics, Department of Physics, The University of Oxford, Keble Road, Oxford, OX1 3RH, UK}

\author[0000-0003-2828-7720]{Andrew Siemion}
\affiliation{Astrophysics, Department of Physics, The University of Oxford, Keble Road, Oxford, OX1 3RH, UK}
\affiliation{Breakthrough Listen, Astrophysics, Department of Physics, The University of Oxford, Keble Road, Oxford, OX1 3RH, UK}
\affiliation{SETI Institute, 339 Bernardo Ave, Suite 200, Mountain View, CA 94043, USA}
\affiliation{Berkeley SETI Research Centre, University of California, Berkeley, CA 94720, USA}
\affiliation{Department of Physics and Astronomy, University of
Manchester, UK}
\affiliation{University of Malta, Institute of Space Sciences and Astronomy, Msida, MSD2080, Malta}



\begin{abstract}

Despite being operational for only a short time, the Einstein Probe mission, with its large field of view and rapid localisation capabilities, has already significantly advanced the study of rapid variability in the soft X-ray sky. We report the discovery of luminous and variable radio emission from the Einstein Probe fast X-ray transient EP240414a, the second such source with a radio counterpart. The radio emission at $3\,\rm{GHz}$ peaks at $\sim30$ days post explosion and with a spectral luminosity $\sim2\times10^{30}\,\rm{erg}\,\rm{s}^{-1}\,\rm{Hz}^{-1}$, similar to what is seen from long gamma-ray bursts, and distinct from other extra-galactic transients including supernovae and tidal disruption events, although we cannot completely rule out emission from engine driven stellar explosions e.g. the fast blue optical transients. An equipartition analysis of our radio data reveals that an outflow with at least a moderate bulk Lorentz factor ($\Gamma\gtrsim1.6$) with a minimum energy of $\sim10^{48}\,\rm{erg}$ is required to explain our observations. The apparent lack of reported gamma-ray counterpart to EP240414a could suggest that an off-axis or choked jet could be responsible for the radio emission, although a low luminosity gamma-ray burst may have gone undetected. Our observations are consistent with the hypothesis that a significant fraction of extragalactic fast X-ray transients are associated with the deaths of massive stars.

\end{abstract}



\section{Introduction}

Fast X-ray transients (FXTs) are bursts of soft X-ray emission lasting 10s to 1000s of seconds and spanning a wide range of luminosities \citep{quirola2022,vasquez2023}. 
Typically discovered through searches of X-ray telescope data archives (particularly those of Chandra, XMM-Newton, \textit{Swift}, and eROSITA) FXTs were only identified months or years after they occurred, and so prompt multi-wavelength follow-up has to date been sparse \citep[although see][]{soderberg2008,ibrahimzade2024}. 
Based on the soft X-ray emission alone a range of progenitor scenarios have been suggested for FXTs including white dwarf tidal disruption events \citep{jonker2013,glennie2015}, stellar flares \citep{glennie2015}, supernova shock breakout \citep{soderberg2008,novara2020,alp2020}, long GRBs \citep{jonker2013,bauer2017}, and newly born rapidly rotating magnetic neutron stars \citep{xue2019}.
The range of observed FXT luminosities (from $10^{30}\,\rm{erg}\,\rm{s}^{-1}$ for suspected stellar flares to $10^{48}\,\rm{erg}\,\rm{s}^{-1}$ for distant extragalactic events) likely supports a variety of progenitor systems, although this conclusion is made unclear by the difficulty in identifying a host galaxy for the majority of extragalactic FXTs to date \citep[see e.g.][]{eappachen2020,eappachen2023,eappachen2024,vasquez2023}.

The observational paradigm for FXTs has recently undergone a drastic shift with the launch of the Einstein Probe X-ray telescope \citep[EP;][]{yuan2022}. The two instruments on board the EP, the Wide Field X-ray Telescope (WXT) and the Follow-up X-ray Telescope (somewhat unfortunately abbreviated as FXT), allow for both wide sky area monitoring as well as follow-up and localisation regions that range between arcminutes and tens of arcseconds, with FXT candidates reported via public alert streams. These capabilities have allowed for rapid follow-up at optical, radio, and X-ray wavelengths and the detection of multi-wavelength counterparts to EPW20240219aa, EP240305a, EP240309a, and EP240315a. The sources EP240305a and EP240309a are likely Galactic, and have been associated with variability from a Gaia star \citep{liu0305atel} and a cataclysmic variable \citep{buckley2024}, respectively. EPW20240219aa was associated with a sub-threshold event by the Fermi Gamma-ray Burst Monitor and is therefore thought to have been caused by a gamma-ray burst \citep{fletcher2024}. EP240315a was associated with a high redshift ($z\sim4.9$) galaxy and was followed up across the electromagnetic spectrum, being the first EP FXT with a radio and optical counterpart \citep{ep240315a_radio_discovery,gillanders2024}. Analysis of the radio, optical, and X-ray data lead to the conclusion that the event harboured a relativistic jet, and was likely from a long gamma-ray burst \citep{gillanders2024,ricci2024,levan2024,liu2024}. With these events, and the increasing number of new events being discovered discovered and quickly reported by the EP, it is clear that significant progress can now be made in understanding the progenitors to FXTs and their multiwavelength properties.

In this paper we present a radio observing campaign on EP240414a, the second EP extragalactic FXT with a multi-wavelength counterpart. EP240414a was discovered \citep{EPdiscovery} by the WXT at UTC 09:50:12 2024-04-14 (MJD 60414.4099; which we define to be $T_{0}$) with a peak flux of $\sim3\times10^{-9}\,\rm{erg}\,\rm{s}^{-1}\,\rm{cm}^{-2}$ in the $0.5$-$4\,\rm{keV}$ energy band and was seen to fade by 4 dex over the following 2 hours with the EP Follow-up X-ray Telescope \citep{guan2024}. An optical counterpart (AT2024gsa) was discovered within three hours of the FXT detection \citep{aryan_opt} and a redshift of the nearest galaxy was reported to be $z=0.41$ \citep{jonkerredshift}. The rapid and unusual optical lightcurve of the source is reported in \cite{srivastav2024}  with the redshift confirmed to be $z = 0.4018 \pm 0.0010$ \citep[see also][]{vandalen2024}. This implies a peak X-ray luminosity of $\sim10^{48}\,\rm{erg}\,\rm{s}^{-1}$, similar to those seen from GRB afterglows at early times \citep[see e.g.][]{gehrels2009}, at a projected offset of $\sim27\,\rm{kpc}$ from the host \citep[unusually large for a long gamma-ray burst][]{bloom2002,blanchard2016,lyman2017}. We discovered the radio counterpart to EP240414a with the MeerKAT radio interferometer, six days after it was reported as an FXT by the EP, as a $\sim200\,\mu\rm{Jy}$ radio source coincident with the EP error circles \citep{guan2024,EPdiscovery} and coincident with AT2024gsa \citep[][]{radiodiscovery,aryan_opt,srivastav2024,vandalen2024}.

This paper describes our entire radio observing campaign on EP240414a. In Section 2 we present our radio observations from MeerKAT, the Australian Telescope Compact Array, and the Allen Telescope Array, as well as serendipitous observations from large sky area radio surveys. In Sections 3 and 4 we share our results and discuss the emission from EP240414a in the context of other extragalactic transients, and in Section 5 we present our conclusions. Throughout this work we assume a flat cosmology with $H_{0}=70\,\rm{km}\,\rm{s}^{-1}\,\rm{Mpc}^{-1}$, $T_{\rm{CMB}}=2.725\,\rm{K}$, and $\Omega_{\rm m}=0.3$. Distances are calculated using the \texttt{astropy.cosmology} package.

\section{Observations} \label{sec:observations}

\subsection{MeerKAT}
\begin{figure}
	\includegraphics[width=\columnwidth]{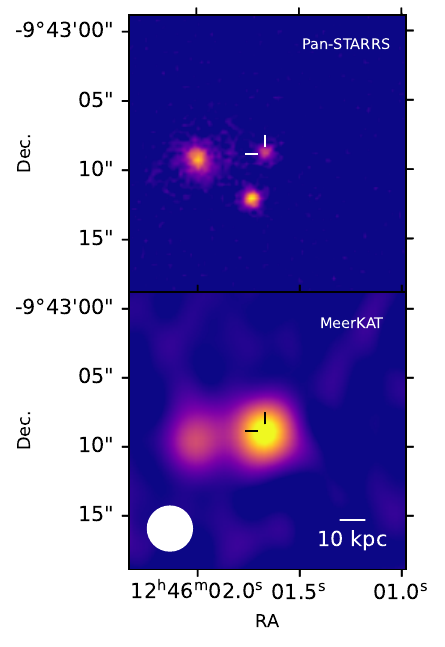}
    \caption{Optical and radio detections of EP240414a/AT2024gsa. \textbf{(top)} A Pan-STARRS i-band image of the field of EP240414a. The source position is marked by a pair of white lines. The diffuse source to the left is the putative host galaxy SDSS J124601.99$-$094309.3 \citep{aryan_opt,jonkerredshift} at a redshift of $z = 0.4018 \pm 0.0010$ (\cite{srivastav2024}, see also \cite{vandalen2024}). \textbf{(bottom)} A subsection of our MeerKAT discovery image of EP240414a with the source position marked by a pair of black lines. The source has a flux density of $227\pm13\,\mu\rm{Jy}$. We detect clear emission from the possible host galaxy of EP240414a. The MeerKAT restoring beam is shown in the bottom left and is $3.2''\times3.2''$ at a position angle of 0 degrees. A scale bar in the bottom right shows $10\,\rm{kpc}$ at $z=0.4$.}
    \label{fig:radio_image}
\end{figure}

The field of EP240414a was observed with the MeerKAT radio telescope under project ID SCI-20230907-JB-01 (PIs Bright \& Carotenuto) where we discovered the $3\,\rm{GHz}$ (using the S4 S-band reciever setup) radio counterpart to EP240414a on 2024 April 20 \citep{radiodiscovery}. The discovery radio image is shown in \Cref{fig:radio_image}, where we detect clear radio emission at the position of EP240414a (RA=12:46:01.669, Dec=-09:43:08.88 from the Transient Name Server where the source is identified as AT2024gsa). We obtained a further two observations of EP240414a with MeerKAT from which source variability can clearly be seen. All of our MeerKAT observations were reduced using the \texttt{oxkat} pipeline \citep{oxkat}, a set of semi-automated scripts for the calibration and imaging of MeerKAT data. \texttt{oxkat} performs phase reference calibration using \texttt{CASA} \citep{casa1,casa2}, self calibration using \texttt{CubiCal} \citep{cubical}, and imaging using \texttt{WSClean} \citep{wsclean}. Images are cleaned using a Briggs robust weighting of -0.3 \citep{briggs1995} and a circular restoring beam is used to create the clean image. The bright and compact source J1239$-$1023 was used as the interleaved phase reference calibrator, while PKS B1934$-$638 was used to set the flux density scale and correct for the bandpass response of the instrument. Typical image RMS values are $5$--$10\,\mu\rm{Jy}\,\rm{beam}^{-1}$. Our MeerKAT observations are summarised in \Cref{tab:radio_observations}.

\subsection{Australia Telescope Compact Array}

We obtained radio observations of EP240414a with the Australia Telescope Compact Array (ATCA) under program CX576 (PI Carotenuto). We first observed EP240414a on 2024 July 2 between 06:10 UT and 10:11 UT. ATCA was in its extended 6D configuration. A second observation, under the same program, was performed on August 23rd 2024 between 04:40 and 08:30 UT, with the telescope in the more compact 1.5A configuration. For both epochs, data were recorded simultaneously at central frequencies of 5.5\,GHz and 9.0\,GHz, with 2\,GHz of bandwidth at each frequency. We used PKS~B1934$-$638 for bandpass and flux density calibration, and J1239--1023 for the complex gain calibration. Data were flagged, calibrated, and imaged using standard procedures within {\tt CASA}. When imaging, we used a Briggs robust parameter of 0 to balance sensitivity and resolution. Our ATCA observations are summarised in \Cref{tab:radio_observations}.

\subsection{Allen Telescope Array}
As part of a larger survey of EP transients we began observing the field of EP240414a on 2024 May 5 with the Allen Telescope Array (ATA; see Farah et al. \textit{in prep.}; Pollak et al. \textit{in prep.}). While multiple observing bands were used \citep[see e.g.][for details]{bright2022} we report only our most constraining limit at a central frequency of $8\,\rm{GHz}$. Observations were reduced with a custom pipeline utilising \texttt{CASA} for calibration and \texttt{wsclean} for imaging. 3C286 was used to set the absolute flux scale and bandpass response of the instrument, while 1246$-$075 was used to correct for the time-dependent complex gains. Our ATA observations are summarised in \Cref{tab:radio_observations}.

\subsection{Archival \& Survey Observations}
We used the Canadian Initiative for Radio Astronomy Data Analysis Image Cutout Provider (\url{https://cirada.ca/services}) to search for observations at the position of EP240414a in large sky area radio surveys. The field of EP240414a was observed with the Karl G. Jansky Very Large Array (VLA) as part of both the VLA Sky Survey \citep[VLASS;][]{lacy2020} and the NRAO VLA Sky Survey \citep[NVSS;][]{condon1998}. Additionally, the Australian Square Kilometer Array Pathfinder Telescope (ASKAP) observed the field as part of Rapid ASKAP Continuum Survey \citep[RACS;][]{mcconnell2020}. For the RACS observation there is no source detected at the position of EP240414a on UTC 2020-10-17, with a 3 sigma upper limit of $\sim950\,\mu\rm{Jy}$ at $885\,\rm{MHz}$. In VLASS epoch 3.2 the upper limit is $\sim420\,\mu\rm{Jy}$ at $3\,\rm{GHz}$ on 2024-07-10 (around 13 weeks after the transient was discovered). The observation as part of NVSS is not constraining compared to the flux density of EP240414a and the limits from RACS and VLASS, and so we do not discuss it further. Radio sky survey observations for EP240414a are summarised in \Cref{tab:radio_observations}.

\begin{table}{}
\centering
\caption{A summary of our radio observations of EP240414a, including measurements from the RACS and VLASS radio sky surveys. $\Delta T$ is given respect to MJD 60414.4099.}
\label{tab:radio_observations}
\begin{tabular}{ccccc}
\hline
Date & $\Delta T$ & Flux Density & Frequency & Facility\\
$[$dd-mm-yy$]$ & [days] & [$\mu$Jy] & [GHz] & \\
\hline
17-10-20 & -1275 & $<905$ & 0.885 & ASKAP \\
20-04-24 & 6.46 & $227\pm13$ & 3 & MeerKAT \\
02-05-24 & 17.79 & $<1050$ & 8 & ATA \\
14-05-24 & 30.48 & $434\pm23$ & 3 & MeerKAT \\
15-06-24 & 62.32 & $304\pm17$& 3 & MeerKAT \\
02-07-24 & 78.92 & $320\pm16$ & 5.5 & ATCA \\
02-07-24 & 78.92 & $240\pm12$ & 9 & ATCA \\
10-07-24 & 86.72 & $<420$ & 3 & VLASS \\
23-08-24 & 130.86 & $113 \pm 12$ & 5.5 & ATCA \\
23-08-24 & 130.86 & $73\pm 8$ & 9 & ATCA \\
\hline
\end{tabular}
\end{table}

\section{Results} \label{sec:results}

\begin{figure}
	\includegraphics[width=\columnwidth]{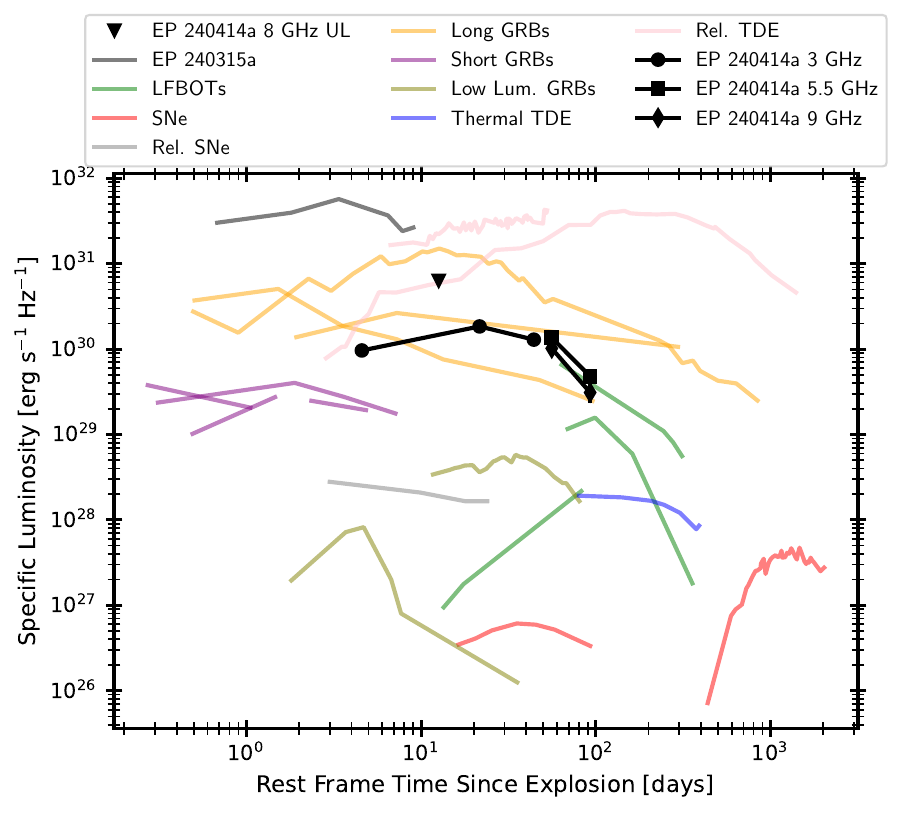}
    \caption{The radio lights curve of EP240414a at 3, 5.5, and $9\,\rm{GHz}$ from MeerKAT and ATCA. The upper limit is from the ATA and is shown with a downward facing arrow. Errors are $1\sigma$ and include both a statistical and absolute calibration uncertainty, but are smaller than the markers. We show other extragalactic transients to help put our data into context. We include (relativistic) supernova, LFBOTs \citep{coppejans2020}, thermal TDEs \citep{alexander2020}, relativistic TDEs \citep{eftekhari2018,rhodes2022}, short gamma-ray bursts \citep{fong2021}, long gamma-ray bursts \citep{berger2002,vanderhorst2008,bright2019}, and low-luminosity gamma-ray bursts \citep{kulkarni1998,soderberg2006}. This figure is based on the one presented in \citet{ho2020}. The specific luminosity has been reduced corrected by a factor of (1+z) to account for the cosmological distances of some sources.}\label{fig:radio_lc}
\end{figure}

Our radio observations of EP240414a are fairly sparse, with three observations at $3\,\rm{GHz}$ with MeerKAT, and two ATCA observations both at $5.5$ and $9\,\rm{GHz}$ (see \Cref{tab:radio_observations} and \Cref{fig:radio_lc}). Our $3\,\rm{GHz}$ light curve shows a clear peak at around $\sim30$ days post explosion in the observer reference frame (or $\sim20$ days in the source rest frame) at $434\pm21\,\mu\rm{Jy}$. While our sampling rate is not high enough to confidently confirm this as the exact peak time at $3\,\rm{GHz}$, we can make the approximation that a spectral peak moved through $3\,\rm{GHz}$ at $30$ days post explosion and therefore take $T_{\rm{pk},3\,\rm{GHz}}=30$ days and $F_{\rm{pk},3\,\rm{GHz}}=434\pm21\,\mu\rm{Jy}$. 

The radio spectral index of our first late time ($\sim80$ days post detection) ATCA measurement is $\alpha=0.58\pm0.15$ where we use the convention $F_{\nu}\propto\nu^{-\alpha}$ and fit a simple power law between the two ATCA frequencies. Assuming this is associated with optically thin synchrotron emission where $\alpha=(p-1)/2$ this implies $p=2.16\pm0.3$ where $p$ is the power-law index of the electron energy distribution ($N(E)\propto E^{-p}$). This indicates that the synchrotron cooling frequency, $\nu_{c}$, is above $9\,\rm{GHz}$ at this epoch, as otherwise a spectral slope of $-p/2$ would be expected and imply an extremely hard electron energy distribution. Our second ATCA epoch suggests a marginal steepening of the spectral index to $0.86\pm0.30$, which would imply $p=2.72\pm0.6$, although the values are statistically consistent. In addition to the spectral index we can make crude estimates for the rise and decay indices of the $3\,\rm{GHz}$ light curves to be $\beta\sim-0.4$ and $\beta\sim0.5$, respectively, for $F_{3\,\rm{GHz}}\propto t^{-\beta}$, although these calculations are obviously sensitive to the exact peak location and assume a constant evolution rate. The late time $3\,\rm{GHz}$ VLASS measurement indicates that no significant re-brightening occurred at late time.

\section{Discussion} \label{sec:disc}

\subsection{Basic considerations}
We begin with a general discussion of the radio properties of EP240414a in the context of other classes of extragalactic radio transients, particularly (relativistic) supernovae, fast blue optical transients, tidal disruption events (both thermal and relativistic), and gamma-ray bursts both long and short (these comparisons can be appreciated graphically by referring to \Cref{fig:radio_lc}). The observed peak 3 GHz specific luminosity of EP240414a is $\gtrsim3\times10^{30}\,\rm{erg}\,\rm{s}^{-1}\,\rm{Hz}^{-1}$ and occurs at $T\sim T_{0}+20\,\rm{d}$ in the rest frame of the explosion. The timing of this peak is significantly earlier than for the radio counterparts to the luminous fast X-ray transients (LFBOTs), which typically peak closer to 100 days post explosion in the GHz range and at luminosities closer to $10^{30}\,\rm{erg}\,\rm{s}^{-1}\,\rm{Hz}^{-1}$ \citep[e.g.][]{margutti2019,ho2019,ho2020,bright2022,coppejans2020}. An exception to this is the FBOT ZTF19abvkwla, which peaked close to $10^{30}\,\rm{erg}\,\rm{s}^{-1}\,\rm{Hz}^{-1}$ and at an uncertain peak time before 100 days post discovery \citep{ho2020}. This can be seen as highest luminosity green line in \Cref{fig:radio_lc} and makes the association between FXTs and FBOTs hard to rule out.

The peak luminosity of EP240414a also exceeds both regular and relativistic core-collapse supernovae by at least an order of magnitude. In a survey of almost 300 supernovae (including a range of different classifications) \citet{bietenholz2021} found a peak specific radio luminosity of $10^{25.5\pm1.6}\,\rm{erg}\,\rm{s}^{-1}\,\rm{Hz}^{-1}$ (including non-detections), with the brightest source in the entire sample having a peak specific luminosity of $\sim10^{29}\,\rm{erg}\,\rm{s}^{-1}\,\rm{Hz}^{-1}$. EP240414a is also more luminous at radio frequencies than all radio-detected superluminous supernovae \citep[SLSN;][]{margutti2023,eftekhari2021}, although radio detections of SLSNe are uncommon, especially at early times.

Tidal disruption events (TDEs) are also know to produce radio emission, either from a relativistic jet or from non-relativistic outflow \citep[e.g.][]{vanvelzen2016,bright2018,eftekhari2018,alexander2020,rhodes2022}. These are termed `relativistic' and `thermal' TDEs, respectively. Relativistic TDEs are some of the most luminous radio transients known, with peak specific luminosities exceeding $\sim10^{32}\,\rm{erg}\,\rm{s}^{-1}\,\rm{Hz}^{-1}$ and peaking at 100s of days post explosion (although the sample of such events is still relatively small). Thermal TDEs peak on similarly long timescales, but at much lower specific luminosities of $\sim10^{29}\,\rm{erg}\,\rm{s}^{-1}\,\rm{Hz}^{-1}$. These peak luminosities and timescales are incompatible with those seen from EP240414a. Any potential TDE association for EP240414a is further disfavoured due to the significant projected offset ($\sim27\,\rm{kpc}$) from the centre of the putative host galaxy \citep[][]{jonkerredshift,levan_sne,srivastav2024,vandalen2024} and see \Cref{fig:radio_image}. 

Finally, we consider EP240414a in the context of gamma-ray bursts (GRBs). Long GRB radio afterglows are caused by two major shocks, the forward and reverse, with radiation from each shock peaking between days and 100s of days post burst, depending on the observing frequency \citep[see e.g.][for examples from well sampled long GRBs with two clear shock components]{laskar2013,vanderhorst2014,bright2023}. The peak radio luminosities of long GRBs span a wide range but are typically between $10^{30}$ and $\sim10^{32}\,\rm{erg}\,\rm{s}^{-1}\,\rm{Hz}^{-1}$ at $\sim10$ days post discovery \citep[see e.g. figure 6 of][]{chandra2012}. The peak luminosity and timescale of EP240414a are both consistent with originating from a long GRB afterglow. The most significant evidence against a GRB progenitor is the location of EP240414a within its host galaxy. Numerous studies \citep{bloom2002,blanchard2016,lyman2017} have shown that long GRBs typically occur close to the centre of their host galaxy, with 80\% of the sample presented in \citet{lyman2017} located within $\sim3\,\rm{kpc}$. In the sample presented in \citealt{blanchard2016} there were no offsets above $20\,\rm{kpc}$ in their sample of $\sim100$ LGBRs. In fact, the galactic offset of EP240414a ($\sim27\,\rm{kpc}$) is much more consistent with those seen for short Gamma-ray bursts, with \citet{fong2022} finding that a significant fraction of their sample fell outside of a $10\,\rm{kpc}$ radius and the most distant above $50\,\rm{kpc}$ in separation. However, EP240414a has been associated with a type Ic broad line supernovae (\citealt{levan_sne, vandalen2024}), a supernova class solidly associated with long gamma-ray bursts \citep[see][for a review]{hjorth2012}. The offset discrepancy was also commented on by \citealt{vandalen2024}, where they note that luminous FBOTs can have host offsets comparable to EP240414a and that there is only weak star formation at the location of EP240414a in its host galaxy. While luminosity rise-time arguments are not conclusive in determining the progenitor of a source, based on the arguments above we suggest that EP240414a is likely a long gamma-ray burst that occurred in an unusual location in its host galaxy. The lack of reported gamma-ray emission suggests that this GRB could have had a low isotropic gamma-ray luminosity, or the radio emission could have been seen from off-axis. For example, the lack of detection by Konus-Wind suggests that the isotropic equivalent gamma-ray energy was below $\sim10^{51}\,\rm{erg}$, based on the sample presented in \citet{tsvetkova2017} (their figure 8). This energy is at the low end of the distribution for long GRBs, but not unusually so \citep[see figure 1 in][]{perley2014}. Following up sources detected with the Einstein Probe could help elucidate the distinction, or lack thereof \citep{nakar2015}, between low luminosity LGRBs and regular LGRBs, potentially driven by differences in circumburst material or viewing angle \citep{liang2007,salafia2016}. 

\subsection{Comparison with FXT EP240315a}

The first FXT with a radio counterpart, EP240315a, was associated with a high redshift ($z=4.9)$ gamma-ray burst \citep{gillanders2024,levan2024,liu2024}. The specific radio luminosity of EP240315a (also at $3\,\rm{GHz}$) was around one order of magnitude larger than the (observed) peak for EP240414a and occurred an order of magnitude earlier (although at $5.5\,\rm{GHz}$, not $3\,\rm{GHz}$). While the light curve properties of EP240315a and EP240414a are not similar, both sources are consistent with the diversity in radio light curves seen from long GRBs \citep[see e.g.][]{chandra2012}. 

\subsection{Minimum Energy Constraints}
\subsubsection{Low Bulk Velocity}
We start by considering our radio observations of EP240414a in the context of minimum energy arguments under the assumption that the emitting region is not moving relativistically. It can be shown \citep[e.g.][]{readhead1977} that the total energy of a synchrotron emitting region, which is the combined energy in the magnetic field and electrons, has a strong minimum as a function of magnetic field strength. This minimum is known as the equipartition energy as it occurs close to (but not exactly at) the magnetic field strength where the energy in the magnetic field and electrons are equal. If the size of the emitting region is known then the magnetic field and therefore the minimum energy can be calculated. 

For the majority of extragalactic transients the size of the emitting region is unknown and so an additional constraint relating the magnetic field and the size of the emitting region is required. Such a condition exists if the emitting region is seen to be self-absorbed to synchrotron radiation at a given frequency, which allows for the minimum energy, magnetic field, and size (and therefore velocity) to be uniquely constrained. Focusing on the energy and size, following \citet{matsumoto2023} we have 

\begin{subequations}
\begin{equation}
E_{\rm{eq,N}}=6.2\times10^{49}d_{L,28}^{40/17}F_{\nu,\rm{mJy}}^{20/17}\nu_{\rm{10}}^{-1}(1+z)^{-37/17}\,\rm{erg}
\end{equation}    
\begin{equation}
R_{\rm{eq,N}}=1.9\times10^{17}d_{L,28}^{16/17}F_{\nu,\rm{mJy}}^{8/17}\nu_{\rm{10}}^{-1}(1+z)^{-25/17}\,\rm{cm}
\end{equation}
\end{subequations}

\noindent
for a completely filled emitting region in the Newtonian limit. The distance to the source ($d_{L,28}$) is given in units of $10^{28}\,\rm{cm}$, the flux density ($F_{\nu,\rm{mJy}}$) in mJy, the frequency ($\nu$) in units of $10\,\rm{GHz}$, and $z$ is the redshift of the source. Using the peak of our $3\,\rm{GHz}$ light curve this implies $E_{\rm{eq}}=1.5\times10^{49}\,\rm{erg}$ and $R_{\rm{eq}}=1.8\times10^{17}\,\rm{cm}$. At 20 days post explosion (in the source rest frame) this size implies an expansion velocity $\beta_{\rm{eq,N}}\approx3.3c$. Note that while \citet{matsumoto2023} assumes a different geometry to \citet{fender2019} (a conical outflow and an expanding sphere, respectively) the derived radii and energies agree to within a factor of order unity. This result implies that the source has a significant, likely at least mildly relativistic, expansion velocity.

\subsubsection{Relativistic Considerations}
In the relativistic regime, minimum energy arguments become more complex due to the dependence of parameters on the bulk Lorentz factor ($\Gamma$) and the angle to the line of sight ($\theta$) to the observer. This manifests through the relativistic Doppler factor $\delta=\Gamma^{-1}(1-\beta\cos\theta)^{-1}$ (as well as through the filling factors, which have a dependence on $\Gamma$). The addition of these parameters means that no global minimum exists for the energy as a function of radius, $\Gamma$, and $\theta$, and instead the total energy must be left as

\begin{equation}\label{eq:1}
\begin{split}
e(R, \theta, \Gamma) &= E/E_{\rm eq,N} \\ &=\frac{\Gamma}{\delta_{\rm D}^{\frac{43}{17}}}\Bigg[\frac{11}{17}\Bigg(\frac{r}{\Gamma\delta_{\rm D}^{-\frac{7}{17}}}\Bigg)^{-6}+\frac{6}{17}\Bigg(\frac{r}{\Gamma\delta_{D}^{-\frac{7}{17}}}\Bigg)^{11}\Bigg]
\end{split}
\end{equation}

\noindent
following \citet{matsumoto2023}, where $r=R/R_{\rm eq,N}$. This reduces to the Newtonian case for $\Gamma,\delta_{\rm D}\rightarrow1$. When fixing one of the parameters ($R$, $\theta$, or $\Gamma$) a family of relativistic minimum energy solutions exists for the pair of remaining parameters. To make progress, we must include an additional relationship between $r=R/R_{\rm eq,N}$, $\Gamma$, and $\theta$, which in both \citet{duran2013} and \citet{matsumoto2023} is $r=(\beta/\beta_{\rm eq,N})\Gamma\delta_{\rm D}$, where

\begin{equation}\label{eq:2}
\begin{split}
\beta_{\rm eq,N}&=\frac{(1+z)R_{\rm eq,N}}{ct}\\
&\approx0.73\Bigg[\frac{F_{p, \rm mJy}^{\frac{8}{17}}d_{L,28}^{\frac{16}{17}}\eta^{\frac{35}{51}}}{\nu_{p,10}(1+z)^{\frac{8}{17}}}\Big(\frac{t}{100\,\rm{days}}\Big)^{-1}\Bigg]f_{A}^{-\frac{7}{17}}f_{V}^{-\frac{1}{17}}
\end{split}
\end{equation}

\noindent
is the Newtonian expansion velocity which can be calculated directly from the peak in the radio spectrum, as done in the previous section \citep[see also][]{duran2013,fender2019}. The value of $\eta$ depends on the ordering of $\nu_{\rm m}$ and $\nu_{\rm sa}$ (the frequencies corresponding to emission from the minimum electron energy in the distribution, and the self absorption frequency, respectively) and is equal to 1 if $\nu_{\rm sa}>\nu_{\rm m}$ (i.e. the spectral peak is due to self absorption) and equal to $\nu_{\rm m}/\nu_{\rm sa}$ if $\nu_{\rm sa}<\nu_m$ (i.e. the spectral peak is due to the minimum electron energy). Area and volume filling factors are given by $f_{A}$ and $f_{V}$, respectively. Recasting \Cref{eq:2} into units more appropriate for Galactic transients reproduces equation (28) from \citet{fender2019} with differences of order unity due to the different assumed geometries.

\begin{figure}
	\includegraphics[width=\columnwidth]{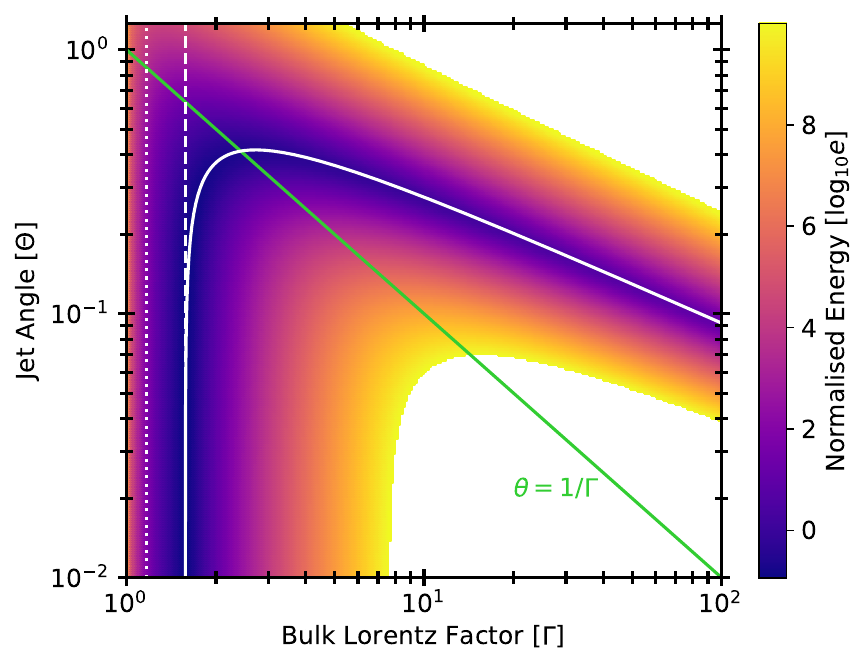}
    \caption{The relativistic energy as a function of bulk Lorentz factor and jet angle for $\beta_{\rm{eq,N}}=3.3$, as is appropriate for EP240414a. The green line divides on- and off-axis solutions to the left and right, respectively. We only show solutions below $10^{10}$ times the minimum energy in the Newtonian limit. The solid white line marks the minimum energy for a given angle and Lorentz factor. The dashed white line shows the asymptotic limit of $\Gamma$ which satisfies the minimum energy condition for $\theta\rightarrow1$ according to \Cref{eq:3}. The dotted white line shows the solution from \citet{matsumoto2023} which breaks down for moderate values of $\beta_{\rm eq,N}$.} 
    \label{fig:min_en}
\end{figure}

\begin{figure}
\includegraphics[width=\columnwidth]{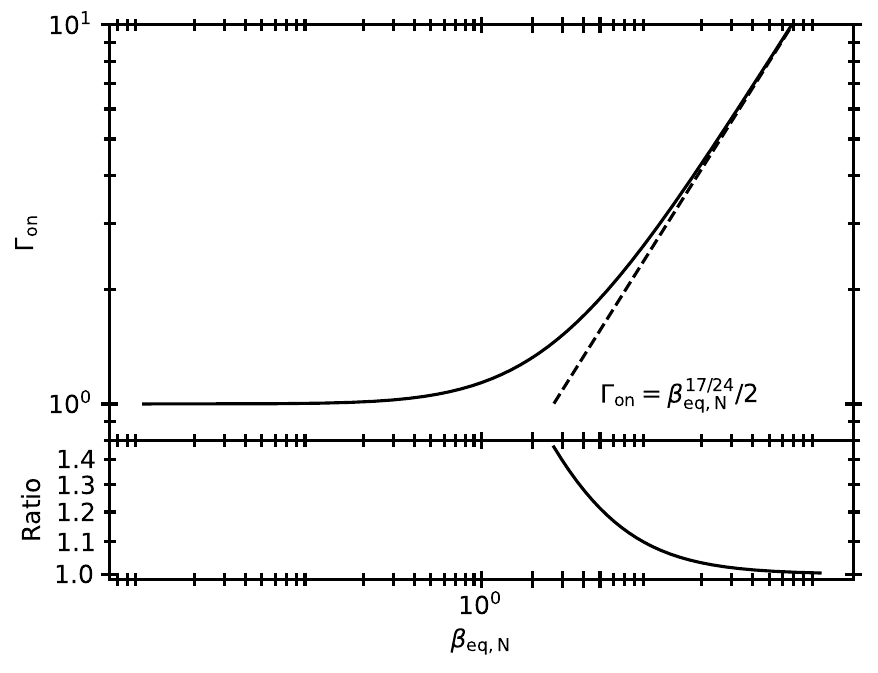}
    \caption{Solution for the minimum bulk Lorentz factor for \Cref{eq:3} and the one given in \citet{matsumoto2023}, which becomes significantly incorrect for $\beta_{\rm eq,N}\lesssim15$ at which point the ratio becomes larger than 1.05. At $\beta_{\rm eq,N}\lesssim9$ the ratio surpasses 1.1.} 
    \label{fig:gamma_on}
\end{figure}

Assuming that the peak we measure in the MeerKAT light curve is due to synchrotron self absorption ($\eta=1$) and taking $f_{A}=f_{V}=1$ we derive $\beta_{\rm eq,N}\approx3.3$ (as before). Recasting \Cref{eq:1} in terms of just $\Gamma$ and $\theta$ it can be seen that there are a family of minimum energy solutions for $(\Gamma,\theta)$ for a given $\beta_{\rm{eq,N}}$, where both on- and off-axis jet solutions exist for different jet angles, depending on the Lorentz factor as $\theta=1/\Gamma$. We note that the condition $\beta_{\rm{eq,N}}\approx3.3$ is robust to the exact values of $\eta$, $f_{A}$, and $f_{V}$ as $\eta\geq1$ and $f_{A},f_{V}\leq1$. We show the minimum energy parameter space for $\beta_{\rm eq,N}=3.3$ in \Cref{fig:min_en}. Considering only on-axis solutions \Cref{fig:min_en} demonstrates that a modest bulk Lorentz $\Gamma\gtrsim1.6$ satisfies the relativistic minimum energy condition for $\theta\lesssim0.4\approx20^{\circ}$. Formally, \cite{matsumoto2023} identify an approximation for the minimum on-axis Lorentz factor in the case that $\theta\ll1$ and $\Gamma\gg1$ as $\Gamma_{\rm on}\approx\beta_{\rm eq,N}^{17/24}/2$, which is not appropriate for EP240414a as it significantly under-predicts the on-axis Lorentz factor to be $\Gamma_{\rm on}\approx1.2$ (see \Cref{fig:min_en}). A more general solution for the jet speed corresponding to the relativistic minimum energy in the on axis case exists as

\begin{equation}\label{eq:3}
\beta_{\rm eq,N}^{\frac{17}{12}}=\beta_{\rm on}^{\frac{17}{12}}\Bigg[\frac{1+\beta_{\rm on}}{1-\beta_{\rm on}}\Bigg]
\end{equation}

\noindent
which can be solved for $\beta_{\rm on}$, and therefore $\Gamma_{\rm on}$, numerically. We demonstrate the solution to \Cref{eq:3} in \Cref{fig:min_en} for $\beta_{\rm eq,N}=3.3$. Recasting \Cref{eq:3} in terms of $\Gamma_{\rm on}$ and taking $\Gamma\gg1$ reproduces the result given in \citet{matsumoto2023} (their equation (34)). We show $\Gamma_{\rm on}$ calculated via these two different methods in \Cref{fig:gamma_on} and demonstrate that the approximation in \citet{matsumoto2023} is inappropriate for $\beta_{\rm eq,N}\lesssim9$ ($\beta_{\rm eq,N}\lesssim15$) at the 10\% (5\%) level.

\begin{figure}
	\includegraphics[width=\columnwidth]{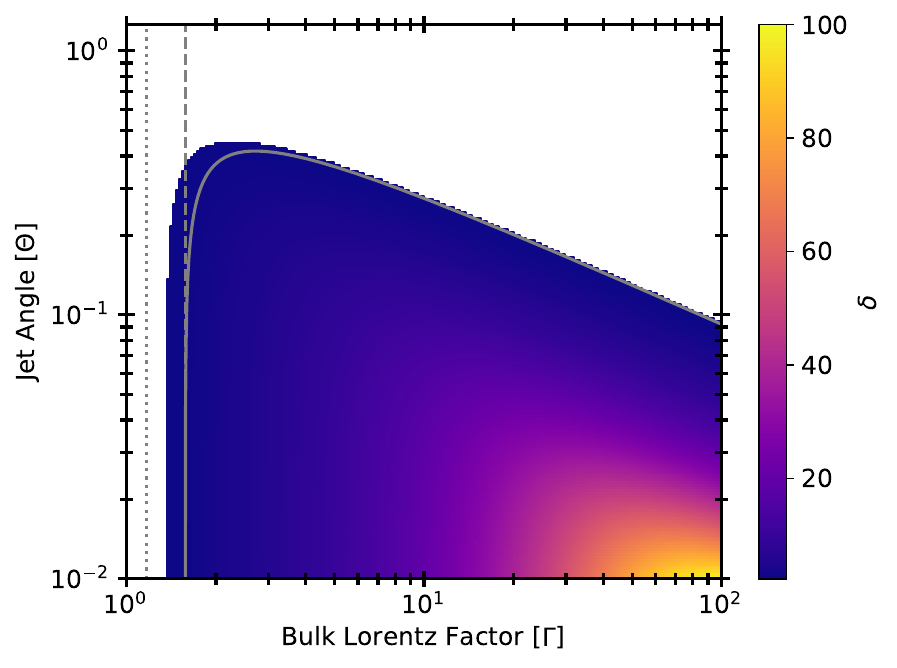}
    \caption{The Doppler factor as a function of $\Gamma$ and $\theta$. The region of the plot coloured white violates the condition $\delta^{49/34}>3.3$, and therefore the boundary is defined as $\delta_{D}^{49/34}=3.3$. It can be seen that the minimum energy locus is approximated by the boundary, and they converge as $\beta\rightarrow1$ (see figure 29 in \citealt{matsumoto2023}, and \citealt{fender2019}). The grey lines are the same as the whites ones in \Cref{fig:min_en}.} 
    \label{fig:min_dop}
\end{figure}

For our measured $\beta_{\rm{eq},\rm{N}}=3.3$ we have that $\beta_{\rm{on}}\approx0.8$ or $\Gamma_{\rm{on}}\approx1.6$. We can compare this with simply considering the Doppler factor required for an apparent velocity of $3.3c$ to appear sub-relativistic. From \citep{fender2019} this requires $\delta_{D}^{49/34}\gtrsim3.3$ implying a velocity above $0.7c$ and a jet angle below $\simeq25^{\circ}$. This is similar to the result derived using the methodology outlined above. Further, the condition $\delta_{D}^{49/34}=\beta_{\rm{eq,N}}$ provides a reasonable approximation to the minimum energy as a function of $\Gamma$ and $\theta$ (the solid white line in \Cref{fig:min_en}) and approaches it in the limit $\beta\rightarrow1$. We demonstrate this in \Cref{fig:min_dop} for $\beta_{\rm{eq,N}}=3.3$.

Our results imply that the emitting region responsible for the radio emission from EP240414a was expanding at least moderately relativistically at 30 days post explosion. The Newtonian equipartition energy ($E_{\rm{eq},\rm{N}}\approx1.5\times10^{49}\,\rm{erg}$) is reduced by a factor of $\sim10$ for the on-axis relativistic case (e.g. $e_{\rm{on}}=0.1$). The potentially large Lorentz factors and uncertain geometries of GRB jets make comparison challenging, for example if the opening angle of the jet $\theta_{j}>1/\Gamma$ then the energy is underestimated by a factor or $4\Gamma^{2}(1-\cos\theta_{j})$.

\begin{figure}
\includegraphics[width=\columnwidth]{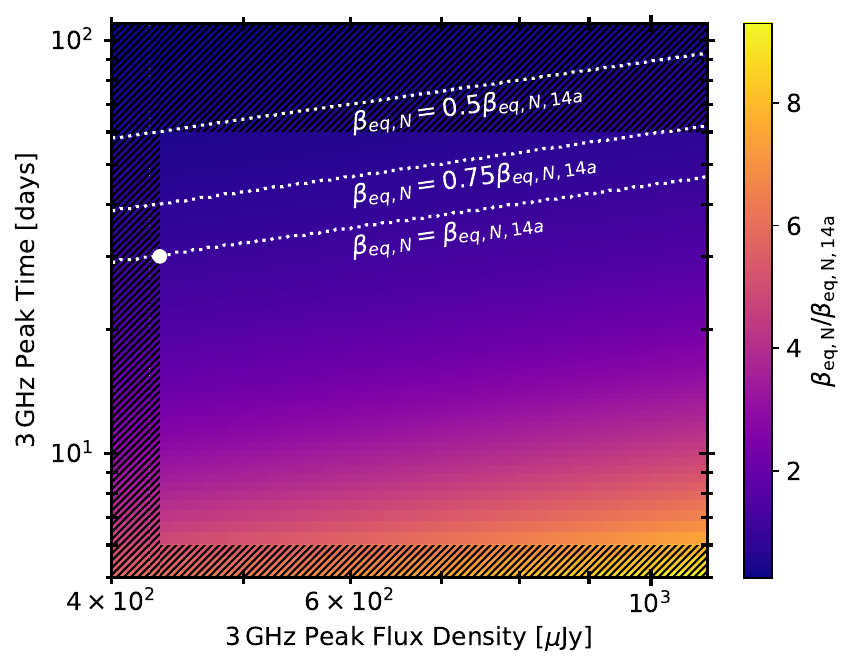}
    \caption{Measured $\beta_{\rm eq,N}$ as a function of an uncertain peak flux density and time expressed as a ratio to $\beta_{\rm eq,N}$ assuming a peak at $\sim30$ days post discovery (giving $\beta_{\rm eq,N,14a}\approx3.3$) which is marked on the plot by a white circle. Hatched areas mark regions ruled out by our observations. We show three white dotted lines to demonstrate $\beta_{\rm eq,N}=(0.5,0.75,1)\times\beta_{\rm eq,N,14a}$. A majority of the allowed parameter space would imply $\beta_{\rm eq,N}$ higher than the one we measure from our assumed peak, and no solutions exist for $\beta_{\rm eq}\lesssim1.7$.} 
    \label{fig:peak_unc}
\end{figure}

Due to our sparse temporal sampling there is significant uncertainty on the peak flux density and frequency at $3\,\rm{GHz}$. We demonstrate in \Cref{fig:peak_unc} that only a small region of the peak flux density and peak time parameter space would have produced a $\beta_{\rm eq,N}$ less than the one we inferred, and no solutions exists for $\beta_{\rm eq,N}\lesssim2$ (or $\Gamma_{\rm on}\gtrsim1.3$). While the upper bound of the peak flux density is not constrained, our ATA observations indicate that EP240414a was likely not more than a few mJy at peak.

\section{Conclusions} \label{sec:conc}

We present radio observations of EP240414a, the second FXT with a radio counterpart. The rise time and peak luminosity of EP240414a distinguish it from regular supernovae and tidal disruption events but are commonly seen from gamma-ray bursts and perhaps from the most luminous fast blue optical transients. Due to the association of EP240414a with a Ic broad line supernova \citep{vandalen2024} a short gamma-ray burst is disfavoured despite the large spatial offset from the host galaxy. Based on minimum energy arguments and the presence of a turn-over in our $3\,\rm{GHz}$ light curve of EP2402414a we place approximate limits on the bulk Lorentz factor of the outflow of $\Gamma\gtrsim1.6$ with an energy $\sim10^{48}\,\rm{erg}$ in the on-axis case. Additionally, we extend the results presented in \citet{matsumoto2023} to show that a general minimum energy solution for on-axis ($\theta\rightarrow0$) sources exists for any measured apparent source velocity, without the assumption of a large bulk Lorentz factor (\Cref{eq:3}). Our observations suggest that at least a moderately relativistic outflow was present in EP240414a and that the progenitor and engine was likely a collapsar, such as those that produce the known population of gamma ray bursts. The absence of a reported gamma-ray counterpart to EP240414a suggests that it could be a GRB with a relatively low isotropic equivalent gamma-ray energy (below $\sim10^{51}\,\rm{erg}$), or that the radio emission is driven by a mildly off-axis jet \citep[as as been suggested by e.g.][]{ibrahimzade2024,wirchern2024} or a cocoon associated with a choked jet \citep[e.g.][]{bromberg2012}. Such scenarios have been suggested for relativistic supernovae such as SN2009bb \citep[e.g.][]{soderberg2010}, however EP240414a is significantly more luminous than such events. Based on the observations presented in this work, \citet{gillanders2024}, and \citet{ricci2024}, we suggest that a significant fraction of EP FXTs are associated with the collapse of massive stars and their afterglow.

In the approximately half a year that the Einstein Probe has been operating it has already redefined the study of the transient X-ray sky, allowing for prompt multi-wavelength follow-up of FXTs. During this time EP FXTs have been firmly associated with flaring stars, cataclysmic variables, and, in the cases of EP240315a and EP240414a, the detected radio emission indicates the production of at least moderately relativistic outflows. Over the coming years the number of FXTs with multi-wavelength counterparts is due to increase dramatically and only through further dedicated observing campaigns will the range of FXT progenitors be elucidated.

\section*{ACKNOWLEDGMENTS}
%%% MeerKAT
The MeerKAT telescope is operated by the South African Radio Astronomy Observatory, which is a facility of the National Research Foundation, an agency of the Department of Science and Innovation.

%%% S-band
This work has made use of the “MPIfR S-band receiver system” designed, constructed and maintained by funding of the MPI f{\"u}r Radioastronomy and the Max-Planck-Society.

%%% ATCA
The Australia Telescope Compact Array is part of the Australia Telescope National Facility (\url{https://ror.org/05qajvd42}) which is funded by the Australian Government for operation as a National Facility managed by CSIRO. We acknowledge the Gomeroi people as the Traditional Owners of the Observatory site.

%%% ATA
The Allen Telescope Array refurbishment program and its ongoing operations are being substantially funded through the Franklin Antonio Bequest. Additional contributions from Frank Levinson, Greg Papadopoulos, the Breakthrough Listen Initiative and other private donors have been instrumental in the renewal of the ATA. Breakthrough Listen is managed by the Breakthrough Initiatives, sponsored by the Breakthrough Prize Foundation. The Paul G. Allen Family Foundation provided major support for the design and construction of the ATA, alongside contributions from Nathan Myhrvold, Xilinx Corporation, Sun Microsystems, and other private donors. The ATA has also been supported by contributions from the US Naval Observatory and the US National Science Foundation

%%% VLA
The National Radio Astronomy Observatory is a facility of the National Science Foundation operated under cooperative agreement by Associated Universities, Inc.

%%% RACS
This scientific work uses data obtained from Inyarrimanha Ilgari Bundara / the Murchison Radio-astronomy Observatory. We acknowledge the Wajarri Yamaji People as the Traditional Owners and native title holders of the Observatory site. CSIRO’s ASKAP radio telescope is part of the Australia Telescope National Facility (\url{https://ror.org/05qajvd42}). Operation of ASKAP is funded by the Australian Government with support from the National Collaborative Research Infrastructure Strategy. ASKAP uses the resources of the Pawsey Supercomputing Research Centre. Establishment of ASKAP, Inyarrimanha Ilgari Bundara, the CSIRO Murchison Radio-astronomy Observatory and the Pawsey Supercomputing Research Centre are initiatives of the Australian Government, with support from the Government of Western Australia and the Science and Industry Endowment Fund. This paper includes archived data obtained through the CSIRO ASKAP Science Data Archive, CASDA (\url{https://data.csiro.au}).

%%%
JM acknowledges funding from a Royal Society University Research Fellowship.
%%%
AM acknowledges support from a Leverhulme Trust International Professorship grant [number LIP-202-014].
%%%
We thank Anna Ho for making the code associated with \citet{ho2020} publicly available.
%%%
FC acknowledges support from the Royal Society through the Newton International Fellowship programme (NIF/R1/211296).
%%%
SJS acknowledges funding from STFC Grants ST/Y001605/1, ST/X006506/1, ST/T000198/1, a Royal Society Research Professorship and the Hintze Charitable Foundation.


\vspace{5mm}
\facilities{ASKAP, ATA, ATCA, MeerKAT, VLA}


\software{astropy \citep{astropy},  
          CASA \citep{casa1,casa2},
          CubiCal \citep{cubical},
          WSClean \citep{wsclean}
          }



\bibliography{EP240414aRadioApJ.bib}{}
\bibliographystyle{aasjournal}



\end{document}